\documentclass[twocolumn,aps,prl,superscriptaddress,showpacs]{revtex4}
\usepackage{amsmath,bm}%
\usepackage{graphicx}%
\usepackage{epsf,epsfig,latexsym}

\begin{document}

\title{ Evidence of Critical Behavior in the Disassembly of Nuclei with A $\sim $36}

\author{Y. G. Ma}
\affiliation{Cyclotron Institute, Texas A\&M University, College Station, Texas}
\affiliation{Shanghai Institute of Nuclear Research,
Chinese Academy of Sciences, Shanghai 201800, China}
\author{R. Wada}
\author{K. Hagel}
\author{J. Wang}
\affiliation{Cyclotron Institute, Texas A\&M University, College Station, Texas}
\author{T. Keutgen}
\affiliation{UCL, Louvain-la-Neuve, Belgium}
\author{Z. Majka}
\affiliation{Jagiellonian University, Krakow, Poland}

\author{M. Murray}
\author{L. Qin}
\author{P. Smith}
\author{J. B. Natowitz}
\affiliation{Cyclotron Institute, Texas A\&M University, College Station, Texas}
\author{R. Alfaro}
\affiliation{Instituto de Fisica, UNAM, Mexico City, Mexico}
\author{J. Cibor}
\affiliation{Institute of Nuclear Physics, Krakow, Poland}
\author{M. Cinausero}
\affiliation{INFN, Laboratori Nazionali di Legnaro, I-35020 Legnaro, Italy  }
\author{Y. El Masri}
\affiliation{UCL, Louvain-la-Neuve, Belgium}
\author{D. Fabris}
\affiliation{INFN and Dipartimento di Fisica dell'Universit\'a di Padova, I-35131 Padova, Italy  }
\author{E. Fioretto}
\affiliation{INFN, Laboratori Nazionali di Legnaro, I-35020 Legnaro, Italy }
\author{A. Keksis}
\affiliation{Cyclotron Institute, Texas A\&M University, College Station, Texas}

\author{ M. Lunardon}
\affiliation{INFN and Dipartimento di Fisica dell'Universit\'a di Padova, I-35131 Padova, Italy  }
\author{A. Makeev}
\affiliation{Cyclotron Institute, Texas A\&M University, College Station, Texas}
\author{N. Marie}
\affiliation{LPC, IN2P3-CNRS, ISMRA et Universit\'e, F-14050 Caen Cedex, France}
\author{ E. Martin}
\affiliation{Cyclotron Institute, Texas A\&M University, College Station, Texas}
\author{A. Martinez-Davalos}
\author{A. Menchaca-Rocha}
\affiliation{Instituto de Fisica, UNAM, Mexico City, Mexico}
\author{G. Nebbia}
\affiliation{INFN and Dipartimento di Fisica dell'Universit\'a di Padova, I-35131 Padova, Italy }
\author{G. Prete}
\affiliation{INFN, Laboratori Nazionali di Legnaro, I-35020 Legnaro, Italy  }

\author{V. Rizzi}
\affiliation{INFN and Dipartimento di Fisica dell'Universit\'a di Padova, I-35131 Padova, Italy   }
\author{A. Ruangma}
\author{D. V. Shetty}
\author{G. Souliotis}
\affiliation{Cyclotron Institute, Texas A\&M University, College Station, Texas}
\author{P. Staszel}
\affiliation{Jagiellonian University, Krakow, Poland}
\author{M. Veselsky}
\affiliation{Cyclotron Institute, Texas A\&M University, College
Station, Texas}
\author{G. Viesti}
\affiliation{INFN and Dipartimento di Fisica dell'Universit\'a di Padova, I-35131 Padova, Italy  }
\author{E. M. Winchester}
\author{S. J. Yennello}
\affiliation{Cyclotron Institute, Texas A\&M University, College Station, Texas}

\date{\today}

\begin{abstract}
A wide variety of observables indicate that maximal fluctuations
in the disassembly of hot nuclei with A$\sim$36 occur at an
excitation energy of  5.6$\pm$0.5 MeV/u and temperature of
8.3$\pm$0.5 MeV. Associated with this point of maximal
fluctuations are a number of quantitative indicators of apparent
critical behavior. The associated caloric curve does not appear to
show a flattening such as that
seen for heavier systems. This suggests
that, in contrast to similar signals seen for  
liquid-gas transitions in heavier nuclei, the observed behavior in these 
very light nuclei is associated with a transition much closer to 
the critical point.

\end{abstract}



\maketitle

Most efforts to determine the critical point for the expected
liquid gas-phase transition in finite nucleonic matter have
focused on examinations of the temperature and excitation energy
region where maximal fluctuations in the disassembly
of highly excited nuclei are observed \cite{Campi,Richert,Chomaz_1,
Bonasera}. A variety of
signatures have been employed in the identification of this region
\cite{Campi,Richert,Chomaz_1,Bonasera,Pan,Elliott_1,DAgostino_1, %
Moretto, Elliott_2,Bauer, Natowitz_1} and several 
publications \cite{Elliott_1, DAgostino_1, Moretto, Elliott_2,Bauer,Natowitz_1}
 have reported the observation of
apparent critical behavior. Fisher Droplet Model analyses have
been applied to extract critical parameters \cite{Elliott_1, %
DAgostino_1, Moretto, Elliott_2,Bauer} which are very close to those 
observed for liquid-gas phase transitions in macroscopic 
systems \cite{Fisher}.  Data from the EOS \cite{Hauger_PRC98} and ISiS
\cite{Beaulieu} collaborations have been employed to
construct a co-existence curve for nucleonic matter
\cite{Elliott_2}. Those analyses have proceeded under the
assumption that the point of apparent critical behavior was the
true critical point of the system \cite{Elliott_1,DAgostino_1}. However,
some recent theoretical treatments suggest that apparent signals of critical
behavior may be encountered well away from the actual critical
point \cite{Gulminelli_1,Norenberg} and applications of $\Delta$-scaling
analyses have suggested that the observation of scaling and
power-law mass distributions \cite{Botet_PRL01,Frankland} 
are not sufficient to identify the true critical
point. Recently, several of the present authors
suggested that disassembly for heavier systems occurs within
the co-existence region, at temperatures well below the critical
temperatures \cite{Natowitz_2}. It was further
suggested that the decreasing importance of Coulomb effects makes
the lightest nuclei the most favorable venue for 
investigation of the critical point. 

In this letter 
we report results of an extensive investigation of nuclear disassembly in
nuclei of $A$ $\sim$ 36 excited  to energies as high as
9 MeV/u. We find that the
maximum fluctuations occur at an excitation energy of 5.6$\pm$0.5
MeV and a temperature of 8.3$\pm$0.5 MeV. At this same point, a number of
indications of apparent critical behavior are seen. While this does not guarantee
that the critical point has been reached, we also find that 
the caloric
curve does not exhibit a plateauing at the point of 
maximum fluctuations, in contrast to experimental results 
for heavier systems \cite{Natowitz_1}.  These observations suggest that 
the critical point for these very light nuclei may have been reached.

Using the TAMU NIMROD detector \cite{NIMROD} and beams from
the TAMU K500 Super-conducting Cyclotron, we have probed the
properties of excited quasi projectile-like fragments (QP) produced in the
reactions of 47 MeV/u $^{40}$Ar + $^{58}$Ni. Earlier work on systems at
energies near the Fermi energy have demonstrated the essential
binary nature of such collisions, even at relatively small impact
parameters \cite{Peter}. As a result, these collisions prove to be
very useful in preparing highly excited light nuclei 
\cite{Steckmeyer}.

The charged particle detector array of NIMROD includes 166
individual CsI detectors arranged in 12 rings in polar angles from
$\sim$ $3^\circ$  to $\sim$ $170^\circ$. In these experiments each
forward ring included two Si-Si-CsI telescopes 
and three Si-CsI telescopes to identify intermediate mass
fragments (IMF). The NIMROD neutron ball, which surrounds the
charged particle array, was used to determine the neutron
multiplicities for selected events. The correlation of the charged
particle multiplicity and the neutron multiplicity was used to
select violent collisions. In this work, we developed a new method 
to reconstruct the QP source. We first carry out three source (QP, 
Quasi-Target and a mid-rapidity sources)
fits to the observed energy spectra and angular distributions of the
light charged particles (LCP).  We then employ the
parameters of these fits to control the $\it{event-by-event}$ assignment
of individual LCP to one of the sources using Monte Carlo
sampling techniques. We associate  IMFs with Z$\geq$4 with the 
QP source if they have rapidity $>0.65y_p$, where $y_p$ is the beam rapidity. 
 For the present analysis we have selected reconstructed QP 
events with total charge number $Z_{QP}$ $\geq$ 12 from violent collisions. 
The 28000 reconstructed  QP events comprise $\sim$ $4\%$ of  violent collisions.

The excitation energy distribution was deduced using the energy
balance equation \cite{Cussol} where the kinetic energies of charged
particles, mass excesses and average neutron contributions were
considered. Using results of the source fits we have also evaluated 
(on the average) small corrections for undetected mass and energy 
(mostly as protons), in the sampled events.    
The QP source velocity was determined from momentum conservation 
of all QP detected particles.

Campi plots of the natural log of the largest cluster charge, ln$Z_{max}$,
versus the natural log of the normalized second moment, ln$S_2$, 
($S_2 = \frac{ \sum_{ Z_i \neq Z_{max}} {Z_i}^2 \cdot n_i  }
{\sum_{ Z_i \neq Z_{max}} {Z_i} \cdot n_i }$, 
where $n_i$ is the multiplicity of QP clusters with atomic number $Z_i$), 
are very  instructive in searches for critical
behavior \cite{Campi,DAgostino_1}.  In Figure 1  we present such plots 
for nine selected excitation energy bins. In the low excitation energy region
only the upper (liquid phase) branch exists.
In the range of $E^*/A$ near 5.6  MeV/u, the liquid branch 
and the lower $Z_{max}$ (gas phase) branch are populated essentially equally. 
At the higher $E^*/A$ the gas 
phase branch is strongly dominant.  These results indicate that the region of maximal 
fluctuations signaling a transition between the two phases is to be found near 
5.6 MeV/u. 

\begin{figure}
\includegraphics[scale=0.4]{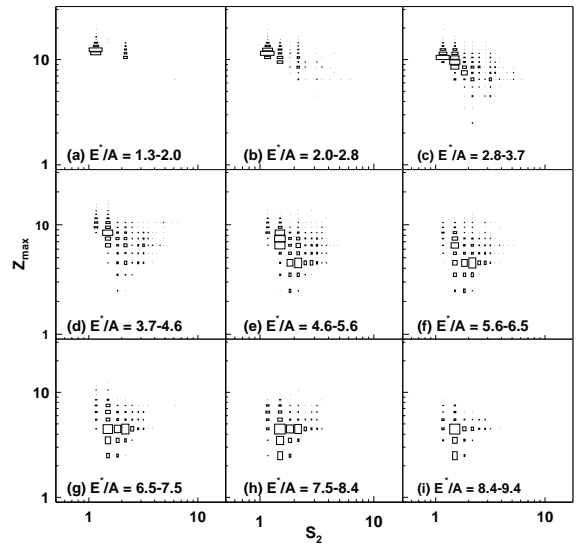}
\caption{\footnotesize Campi plots in different $E^*/A$ windows.}
\label{campi}
\end{figure}
To further explore this region we have investigated other proposed observables 
commonly related to fluctuations and critical behavior.  
Since we wish to compare our results with several model calculations 
which are based upon specification of the initial temperature of the excited 
system, we first turn to a determination of the caloric curve for these A=36 nuclei.   
We have used two different techniques to derive "initial temperatures" from
the observed apparent temperatures. The first consisted of
fitting  the kinetic energy spectra for different LCPs in
9 different bins in $E^*/A$  with  Maxwellian
distributions to obtain the slope temperatures. To derive the initial
temperatures it is necessary that 
the experimentally observed slope temperatures be  corrected for effects due to 
secondary decay. We therefore employed
the measured excitation energy dependence of the multiplicity
for the ejectile under consideration to determine initial temperatures \cite{Hagel,Wada}. 
The second technique employed results of a Quantum Statistical Model 
(QSM) calculation to correct the observed double isotope, H-He, ratio temperatures 
for secondary decay effects \cite{Gulminelli_2,Majka}. For this work we 
employed the QSM model described in reference \cite{Majka}. The first 
technique employing observed spectral slopes  assumes  sequential
evaporation of the ejectiles from a cooling compound nucleus
source \cite{Hagel,Wada} while the second 
assumes simultaneous fragmentation of a reduced density
equilibrated nucleus and subsequent secondary evaporation from the
primary fragments \cite{Gulminelli_2,Majka}. Given that the discussion
above suggests an important transition from liquid to gas dominance at
5.6 MeV/u excitation energy, the first method should not be appropriate
above that energy and the second method is not appropriate below that energy. 
The data points in Figure 2 represent the initial temperature values
determined from cascade corrected slopes (solid squares) and H-He
isotope ratios (solid circles), each determined in its appropriate region 
of applicability.  The two techniques lead to reasonable agreement in the transition region.   
We note that the caloric curve, defined in this manner, exhibits no obvious plateau.
The temperature at 5.6 MeV excitation is 8.3 $\pm$ 0.5 MeV.

\begin{figure}
\vspace{-0.8truein}
\includegraphics[scale=0.35]{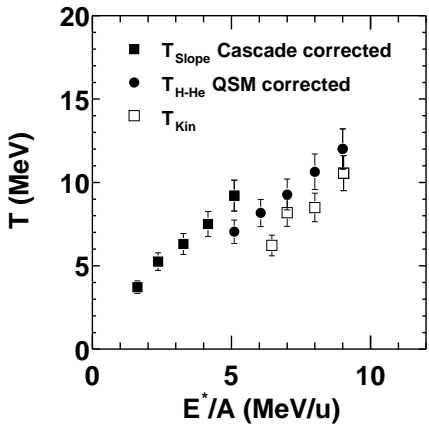}
\vspace{-0.5truein}
\caption{\footnotesize Caloric curve. See text for details. }
\label{caloric}
\end{figure}

If the vapor phase may be characterized as an ideal gas of 
clusters \cite{Fisher}, then, at and above $T$ = 8.3 MeV, this 
should be signaled by a kinetic temperature, 
T$_{\rm{}kin} = \frac{2}{3} {\rm E}^{\rm{}th}_{\rm{}kin}$, 
where E$^{\rm{}th}_{\rm{}kin}$ is the 
Coulomb corrected average kinetic energy of primary fragments. Secondary 
decay effects  make it difficult to test this expectation. However, 
in an inspection of the average kinetic energies for the 
different species observed, we find  that, {\it for 
each $E^*/A$ window}, the average kinetic energy  of 
$^{\rm{}3}$He isotropically emitted in the projectile like frame, is 
higher than those of other species. This together with simple 
model estimates indicates that the $^{\rm{}3}$He spectra are 
the least affected by secondary decay.  Kinetic temperatures for 
$^{\rm{}3}$He, defined as $\frac{2}{3}(\bar{E}_k-B_c)$ where $\bar{E}_k$ 
is the average kinetic energy and $B_c$ is the Coulomb energy, 
are plotted as open squares in Figure 2. Above $T$ = 8.3 MeV  the kinetic temperatures
show a similar 
trend to the chemical temperatures but are 
approximately 1.5 MeV lower. While not perfect this approximate agreement  
provides additional evidence for disassembly of an 
equilibrated system.

Returning to the characterization of the apparent transition at 
$E^*/A$ = 5.6 MeV, we plot in Figure 3(a) the effective 
Fisher-law parameter $\tau_{eff}$ extracteded from power law fits for $Z$ = 2 - 7 
of the QP charge distributions in the different $E^*/A$ windows. 
The results are plotted 
as a function of $T/T_0$, where $T_0$ is  $8.3\pm0.5$ MeV temperature derived for $E^*/A$ 
= 5.6 MeV. As seen in Figure 1, at low $E^*/A$ a large 
residue always remains,  {\it i.e.} the
nucleus is basically in the liquid phase accompanied by some
light particles. When $T/T_0 \sim 1$, 
the charge distribution shows a near power-law
distribution with $\tau_{eff}$ $\sim$ 2.3. This value is close to
the  critical exponent of the liquid gas phase transition
\cite{Fisher}.  As $T$ continues to increase, the charge
distribution becomes steeper which indicates that the system tends
to vaporize.  The observed minimum in $\tau_{eff}$ is rather broad.
For comparison to the experimental data we also display, in Figure 3(a), the variation of 
$\tau_{eff}$ effective predicted by three different models:  
standard statistical sequential decay as modeled 
by the  code GEMINI \cite{Charity}, the isospin dependent 
Lattice Gas Model (LGM) of Das Gupta 
{\it et\ al.} (using A = 36 and Z = 16 particles in a 
 cubic lattice with 64 sites ) \cite{Pan,DasGupta} and Classical 
Molecular Dynamics (CMD) with Coulomb forces \cite{DasGupta}. The last, without Coulomb, has 
been shown to give results essentially the same as the LGM which is known 
to have a liquid gas phase transition. Thus the addition of Coulomb interactions to that 
model is argued to be equivalent to the more difficult task of inclusion of Coulomb 
interactions in the LGM. In order to compare all parameters on a $T/T_0$ scale, 
$T_0$ for the GEMINI calculation is taken as 8.3 MeV. For the LGM and CMD calculations $T_0$ 
is taken as the associated phase transition temperature for A = 36 and Z = 16. 
For the potential employed Ref.\cite{DasGupta} these are, respectively,  
$T_0$ = 5.0 MeV and $T_0$ = 4.5 MeV.
Scaled in this way, the LGM and CMD results exhibit a variation of $\tau_{eff}$  vs 
$T/T_0$ that is very similar to that seen in the experiment while 
the GEMINI calculation shows dramatically different behavior. 

In a LGM investigation of scaling and apparent
critical behavior, Gulminelli {\it et al.} have pointed out that, in
finite systems, the size distribution of the maximum cluster, {\it i.e.}
the liquid, might overlap with the gas cluster distribution in
such a manner as to mimic the critical power law behavior with
$\tau_{eff}$ $\sim$ 2.2 \cite{Gulminelli_1}. They further note,
however, that at that point the scaling laws are satisfied \cite{Gulminelli_1}.
We find that removal of the heaviest cluster from the distribution
leads to distributions which are exponential in nature over the
entire energy range sampled. In Figure 3(b) the resultant
exponential slope parameters, $\lambda_{eff}$, are also plotted
against excitation energy. A minimum is seen in the same region
where $\tau_{eff}$ exhibts a minimum. 
Once again the LGM and CMD models show similar 
behavior while the GEMINI calculation leads to very different predictions.
 
For a further comparison we present  in Figure 3(c) - 3(e) 
experimental and calculated    
results  for (c) the mean normalized second moment, 
$\langle S_2\rangle$, from Figure 1, (d) the 
normalized variance in $Z_{max}/Z_{QP}$ distribution, 
NVZ = $\frac{\sigma^2_{Z_{max}/Z_{QP}}}{\langle Z_{max}/Z_{QP}\rangle}$ \cite{Dorso}, 
(e) $\langle Z_{2max}\rangle$ - the average atomic number 
of the second largest fragment \cite{Sugawa}.
For these additional parameters, which are often used to characterize a region of 
maximum fluctuations, the data show maxima at $T/T_0$ = 1. The LGM and CMD calculations 
show very similar behavior with some difference in absolute value. (Note suppressed zero 
points on y-axes of the plots in Figure 3.) 
The GEMINI results are very different in each case.

\begin{figure}
\vspace{-0.2truein}
\includegraphics[scale=0.4]{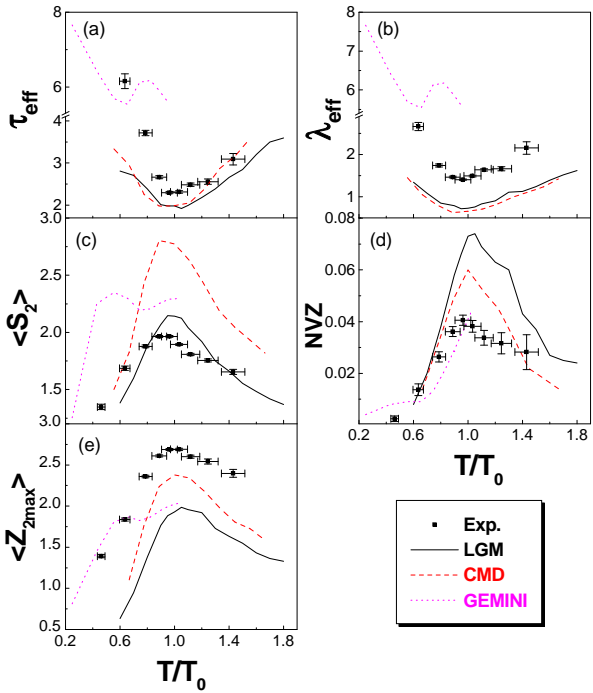}
\vspace{-0.4truein}
\caption{\footnotesize The effective Fisher-law parameter
($\tau_{eff}$) (a), the effective exponential law parameter ($\lambda_{eff}$) (b),
$\langle S_2\rangle$ (c), NVZ fluctuation (d), 
the mean charge number of the second largest fragment $\langle Z_{2max}\rangle$ (e).
Solid squares with the error bars are experimental data and the lines are 
different model calculations as illustrated in
right bottom corner. See details in text.} 
\label{fluctuation}
\end{figure}

To recap, in our measurements for the disassembly  of a  
small nucleus with A$\sim 36$  maximal fluctuations are observed at $5.6\pm0.5$ MeV/u 
excitation energy  and $T$ = $8.3\pm0.5$ MeV. The 
fragment topological structures suggest the onset of a phase change. 
Comparisons with results of LGM and CMD (with Coulomb) calculations 
suggest that the critical point may have been reached. As pointed out 
in the introduction, recent theoretical treatments suggest that apparent 
signals of critical
behavior may be encountered well away from the actual critical
point \cite{Gulminelli_1,Norenberg,Botet_PRL01,Frankland},  
and therefore most of the parameters investigated,   
in Figures 1 and 3 may not be sufficient  by themselves to identify the 
true critical point of the system. 
What differentiates the present work from previous identifications of 
points of critical behavior in nuclei, in addition to the fact that these 
are the lightest nuclei for which a detailed experimental analysis has 
been made, is the comportment of the caloric curve.
For heavier systems the points identified as the critical points appear 
to occur at excitation energies very close to those at which the onset 
of significant flattening in the caloric curves are observed \cite{Natowitz_1}. 
The reason for this flattening is still under discussion. 
It may reflect expansion and/or spinodal decomposition inside the coexistence 
region \cite{Dorso_2,Ono,Norenberg}. In contrast no quasi 
plateau region in the caloric curve is apparent in the present study. 
Further, above the point of maximal fluctuations, the kinetic temperatures 
of $^3He$ ejectiles, thought to represent early emission with little 
contamination from secondary decay, show a similar
 trend to that of the chemical temperatures. 
Also, we believe it is worth noting that the significance of the 
event topology in the $T$ = 8.3 MeV 
region is also indicated in a Zipf's law test (not shown) as proposed in \cite{Ma_1}.
Moreover, the analysis of $\Delta$-scaling of $Z_{max}$ distribution, 
the fluctuation of total kinetic energy of QP and critical exponents  
also support the phase change around $E^*/A \sim 5.6$ MeV \cite{Ma_2}.
Taken together, this body of evidence suggests a phase 
change in an equilibrated system at, or extremely close to, the critical 
point of this mesoscopic system.

This work was supported by the the U.S. Department of Energy and
the Robert A. Welch Foundation. The work of YGM is also supported
by the NSFC under Grant No. 19725521 and the Major State Basic
Research Development Program in China under Contract No. 
G2000077404. R.A, A.M-D and A.M-R wish to thank the partial
support of DGAPA and CONACYT-Mexico. 

{}


\footnotesize
\begin{thebibliography}{}
\bibitem{Campi} X. Campi, J. Phys. A {\bf 19}, L917 (1986); Phys. Lett. 
{\bf B208}, 351 (1988).
\bibitem{Richert} J. Richert and P.Wagner, Phys. Rep. {\bf 350}, 1 (2001). 
\bibitem{Chomaz_1}P. Chomaz, in {\it Proceedings of the INPC 2001 
Conference}, Berkeley, Ca., July 2001.
\bibitem{Bonasera} A. Bonasera {\it et al.} Riv. Del  Nuov. Cim. {\bf 23}, 1 
(2000).
\bibitem{Pan}J. C. Pan {\it et al.}, Phys. Rev. Lett. {\bf 80}, 1182 (1998).
\bibitem{Elliott_1}J. B. Elliott {\it et al.}, Phys. Rev. C {\bf 62}, 
064603 (2000).
\bibitem{DAgostino_1}M. D'Agostino {\it et al.}, Nucl. Phys. {\bf A650}, 
329 (1999).
\bibitem{Moretto} L. G. Moretto {\it et al.}, 
in {\it Proceedings of the INPC 2001 Conference}, Berkeley, Ca., July 2001.
\bibitem{Elliott_2} J. B. Elliott  {\it et al.}, Phys. Rev. Lett. {\bf 88},
 042701 (2002); 
J. B. Elliott {\it et al.}, Phys.\ Rev.\ C {\bf 67}, 024609 (2003). 
\bibitem{Bauer}M. Kleine Berkenbusch {\it et al.}, Phys. Rev. Lett. 
{\bf 88}, 022701 (2002).
\bibitem{Natowitz_1}J. B. Natowitz {\it et al.}, Phys. Rev. C {\bf 65}, 
034618 (2002).
\bibitem{Fisher} M. E. Fisher, Rep. Prog. Phys. {\bf 30}, 615
(1969); Physics {\bf 3}, 255 (1967).
\bibitem{Hauger_PRC98} J. A. Hauger {\it et al.}, Phys. Rev. C {\bf 57}, 
764 (1998); ibid {\bf 62}, 024626(2000).
\bibitem{Beaulieu} L. Beaulieu {\it et al.}, Phys. Rev. Lett. {\bf 84}, 
5971 (2000).
\bibitem{Gulminelli_1} F. Gulminelli {\it et al.}, Phys. Rev. C {\bf 65}, 
051601 (2002); F. Gulminelli and P. Chomaz, Phys. Rev. Lett. {\bf 82}, 
1402 (1999). 
\bibitem{Norenberg} W. Norenberg {\it et al.}, GSI Preprint 2002-3, January, 
2002.
\bibitem{Botet_PRL01} R. Botet {\it et al.}, Phys. Rev. Lett. {\bf 86}, 
3514 (2001).
\bibitem{Frankland} J.D. Frankland {\it et al.}, ArXiv nucl-ex/0201020.
\bibitem{Natowitz_2} J.B. Natowitz {\it et al.}, ArXiv nucl-ex/0206010. 
\bibitem{NIMROD} http://Cyclotron.tamu.edu/nimrod/
\bibitem{Peter}J. P\'eter {\it et al.}, Nucl. Phys. {\bf A593}, 95 (1995).
\bibitem{Steckmeyer}J. C. Steckmeyer {\it et al.}, Phys.\ Rev.\ Lett. 
{\bf 76}, 4895 (1996).
\bibitem{Cussol} D. Cussol {\it et al.}, Nucl. Phys. {\bf A561}, 298 (1993).
\bibitem{Hagel}K. Hagel {\it et al.}, Nucl. Phys. {\bf A486}, 429 (1988).
\bibitem{Wada}R. Wada {\it et al.},  Phys. Rev. C {\bf 39}, 497 (1989).
\bibitem{Gulminelli_2}F. Gulminelli and D. Durand, Nucl. Phys. {\bf A615}, 
117 (1997).  
\bibitem{Majka}Z. Majka {\it et al.}, Phys. Rev. C {\bf 55}, 2991 (1997). 
\bibitem{Charity} R. J. Charity {\it et al.}, Nucl.\ Phys.\ {\bf A483}, 371 (1988).
\bibitem{DasGupta}J. Pan and S. Das Gupta, Phys.\ Rev.\ C {\bf 57}, 1839 (1998).
\bibitem{Dorso}C. O. Dorso {\it et al.}, Phys. Rev. C {\bf 60}, 034606 
(1999).
\bibitem{Sugawa}Y. Sugawa and H. Horiuchi, Prog. Theo. Phys. {\bf 105}, 131 
(2001). 
\bibitem{Dorso_2} A. Strachan and C. O. Dorso, Phys.\ Rev.\ C {\bf 58}, R632 (1998).
\bibitem{Ono} T. Furuta and A. Ono, ArXiv:nucl-th/0305050.
\bibitem{Ma_1}Y. G. Ma, Phys.\ Rev.\ Lett. {\bf 83}, 3617 (1999).
\bibitem{Ma_2} Y. G. Ma (NIMROD Collaboration), talk presentation in HIC03, 
Montreal, Canada, June 25-28, 2003. http://www.physics.mcgill.ca/HIC03/talks.html 
\end{thebibliography}
\end{document}